\documentstyle[preprint,pre,aps,eqsecnum,epsf]{revtex}
\newcommand {\w} {\omega}
\newcommand {\B} {\beta}
\newcommand {\ra} {\rightarrow}
\newcommand {\lc} {\langle}
\newcommand {\rc} {\rangle}

\newcommand {\D} {\Delta}
\newcommand {\sg} {\sigma}
\newcommand {\e} {\epsilon_0}
\newcommand {\nn} {\noindent}
\newcommand {\T} {\tilde}
\newcommand {\ga} {\alpha}
\newcommand {\g} {\gamma}
\begin{document}
\draft


\title{Parametrising yields of nuclear multifragmentation}

\author{C. B. Das, S. Das Gupta and A. Majumder}
\address{
Physics Department, McGill University,
3600 University St., Montr{\'e}al, Qu{\'e}bec \\ Canada H3A 2T8\\ }

\date{ \today } 

\maketitle

\begin{abstract}
We consider a model where, for a finite disintegrating system,
yields of composites can be calculated to arbitrary accuracy.
An analytic answer for yields is also known in the thermodynamic limit.
In the range of temperature and density considered in this work,
the model has a phase transition.  This phase transition is first order.
The analytic expression for yields of composites, in the thermodynamic 
limit, does not conform to the expression 
$<n_a>=a^{-\tau}f(a^{\sigma} (T-T_c))$
where, the usual identification would be that $T_c$ is the critical
temperature and $\tau,\sigma$ are critical exponents.
Nonetheless, for finite systems, we try to fit the yields with
the above expression.  A minimisation procedure is adopted to
get the parameters $T_c,\tau$ and $\sigma$.  While deviations from
the formula are not negligible, one might believe that the deviations
are consistent with the corrections attributable to finite particle
number effects and might then conclude that one has deduced at
least approximately the values of critical parameters.
This exercise thus points to difficulties of trying to extract critical
parameters from data on nuclear disintegration.  An interesting result
is that the value of $T_c$ deduced from the ``best'' fit is very close
to the temperature at which the first order phase transition occurs in the
model.

The yields calculated in this model can also be fitted quite well 
by a parametrisation
derived from a droplet model.
\end{abstract}


\section{INTRODUCTION}

The following parametrisation, often used to fit nuclear 
multifragmentation data, gives an elegant expression for yields of composites:
\begin{eqnarray}
<n_a>=a^{-\tau}f(a^{\sigma}(T-T_c))
\end{eqnarray}
Here $a$ is the mass number of the composite, $T_c$ is the critical 
temperature, $\tau$ is the Fisher exponent \cite{Fisher} and
the expansion is valid in the 
neighbourhood of $T_c$ and for ``large'' $a$.  Variants of the equation
are also used.  For example, in the percolation model one replaces $T-T_c$
by $p-p_c$ where $p_c$ is the value of $p$ at which an
infinite cluster first appears \cite{Stauffer}.  The values of $\tau$
and $\sigma$ for three dimensional percolation model are 2.18 and 0.45
respectively.  In the case of the lattice gas model, also used in a great
deal to study nuclear multifragmentation\cite{Pan1},
Eq (1.1) is valid not only
in the neighbourhood of $T_c$ but along the entire Kertesz line and
extends also to lower than critical density 
\cite{Pan1,Kertesz,Gulminelli1,Pan2}.

There is also another class of models which have been used to fit
multifragmentation data.  Typical of these is the Copenhagen
statistical multifragmentation model \cite{Bo}.  Details vary between 
different versions but the principal
assumption is that, at some larger than normal volume, hot nuclear
matter breaks up into different clusters.  Nuclear interactions
between different clusters are deemed insignificant.  The coulomb
interaction between different composites can be taken into account
at least approximately.  Numerous applications of such models have
been made with impressive success.  The advantage of such models is
that quantum effects such as shell effects can be included.
 
How well does Eq (1.1) work for such models?  This is a relevant question
in view of the fact that there already exist many applications of such 
models to fit actual data \cite{Bo,Majumder,Tsang}.
We take a simplified version
of such models so that exact answers can be obtained (as opposed to
Monte-Carlo sampling which make calculations very long).  The story
that unfolds is quite interesting and is the subject of this paper.
For simplicity, we first consider a model of one kind of particle in the 
next section.  This is then anlysed for parametrisation.
The method of analysis is explained in section 3.  Results are presented 
in section 4.  After looking at fits to models with isotopic spins with
and without the Coulomb force, we return to the model of one kind of
particles again, this time, trying a fit with a droplet model. A comparison
between different models is presented in section 7.
Our conclusions are summarised in section 8.

\section{Model of one kind of particle}
 
Details of the model can be found in \cite{Dasgupta,Bhatta}.  They
are summarised here for completeness.

We first consider just one kind of particles as the thermodynamic properties
for this model were studied in detail \cite{Dasgupta,Bug1,Bug2,Ell}.
We can have monomers or composites of $k$ nucleons.  The composites have
ground state energy $-Wk+\sigma(T)k^{2/3}$.  The first term is the 
volume energy with $W$=16 MeV.  The second term is the surface tension term.
The surface tension 
term is taken to be temperature dependent as in \cite{Bo}:
$\sigma(T)=\sigma_0[(T_c^2-T^2)/(T_C^2+T^2)]^{5/4}$ with $\sigma_0=$18 MeV
and $T_c=18$ MeV.  The internal partition function of a composite of
$k>1$ nucleons is:
\begin{eqnarray}
z_k=\exp[(Wk-\sigma(T)k^{2/3}+T^2k/\epsilon_0)/T]
\end{eqnarray}
where we have used the standard Fermi-gas model for excited states(the
third term on the right hand side of Eq.(2.1)).  For $k=1$ we take $z_1$=1.

We want to construct the canonical partition function of the system
which has total $A$ nucleons in a volume $V$ which is much larger than
the volume of a nucleus of $A$ nucleons in the ground state:$V>V_0$.
We assume when clusters are formed in the volume $V$, they do not
overlap with each other.  Thus the available volume free to the
clusters is less than $V$.  It is given by $V_f=V-V_{exc}$.  We
take $V_{exc}$ to be $V_0=A\times b$ where $b=1/.16 fm^{-3}$.  In 
reality $V_{exc}$ should also depend on multiplicity but this 
complication is ignored here.  The canonical partition function $Q_A(T)$
of $A$ nucleons is then given by
\begin{eqnarray}
Q_A=\sum\prod_{k\geq 1}\frac{\omega_k^{n_k}}{n_k!}
\end{eqnarray}
where $\omega_k$ is the partition function of one composite of $k$
nucleons:
\begin{eqnarray}
\omega_k=\frac{V_f}{h^3}(2\pi mT)^{3/2}k^{3/2}\times z_k
\end{eqnarray} 
and the sum rule must be obeyed:$A=\sum kn_k$.
As noted before \cite{Dasgupta}, the partition function $Q_A$ for
$A$ nucleons can be easily generated in the computer by utilising a
recursion relation.  Starting with $Q_0=1$ one can build all higher
ones using
\begin{eqnarray}
Q_p=\frac{1}{p}\sum_{k=1}^pk\omega_kQ_{p-k}
\end{eqnarray}
The expression for the yield of composites is, of course, of primary
interest.  This is given by
\begin{eqnarray}
<n_k>=\omega_k\frac{Q_{A-k}}{Q_A} \label{canyld}
\end{eqnarray}
Several things are known for this model.  The critical temperature for
the model is the temperature at which the surface tension vanishes
\cite {Bug1} and hence it is at $T=T_c=18$MeV.  The critical volume
is at $V=V_0$.  At temperatures below 18 Mev there is first order phase
transition \cite {Dasgupta,Bug1}. The temperature of phase 
transition depends upon the density.  This temperature was called boiling
temperature in \cite{Dasgupta}.  In the temperature range we are concerned
with in this paper, there is only first order phase transition.  The phase
transition temperature is characterised by a sudden jump in specific
heat.  In finite systems we will take boiling temperature to be the
temperature at which the specific heat maximises.

We will try to fit the yields of Eq.(2.5) by the generic formula
(Eq.(1.1)).  The exact expressions (Eqs.(2.2) to (2.5))
give no clue of a simple 
parametrisation.  Provided $V$ is reasonably bigger than $V_0$ (see
however \cite {Bug1}) and $A$ is also large we can use the grand canonical
results to obtain some insight on simple parametrisation.  This answer
is well-known:
\begin{eqnarray}
<n_k>=\frac{V_f}{h^3}(2\pi mT)^{3/2}k^{3/2}\exp[(\mu(T)+W+T^2/\epsilon_0)k/T
-\sigma(T)k^{2/3}/T] \label{grand}
\end{eqnarray} 
There is no exact correspondence between Eq.(2.6) and Eq.(1.1).  Thus we
may at best hope to get an approximate fit.  How we do it is detailed in
the next section.

\section{The fitting procedure}
Here we follow very elegant methods given in \cite{Gulminelli1}.  
For later use in the text, we will give adequate details.

The quality of fit is given by the smallness of
$\chi^2$.  If the calculated quantity $Y$
is a function of two parameters $Y=Y(a_i,b_i)$ and we are trying to
fit it to a function $g(a_i,b_i,\alpha,\beta,\gamma...)$ then 
$\chi^2=\sum_i(Y(a_i,b_i)-g(a_i,b_i,\alpha,\beta,\gamma....))^2/
\sum_jY(a_j,b_j)^2$.  Variations of this criterion are also possible.
Eq. (1.1) requires us to find ``best'' values for $\tau,\sigma$ and
$T_c$.  This is done in several steps.

(1) At each temperature we find the ``best'' $\tau$
associated with an attempted fit $<n_a>=a^{-\tau}C$ where $C$ is
a constant.  This
follows from Eq.(1.1) but only at $T=T_c$, hence one can argue that
if Eq.(1.1) were exact for yields calculated by Eq.(2.5)
we would get null $\chi^2$ and the correct $\tau$
at $T=T_c$.  This would determine both $T_c$ and $\tau$.  Of course,
null $\chi^2$ is not found since ``exact'' fit is not given by Eq.(1.1)
However we can draw a best ``$\tau$'' {\it vs.} $T$ curve for a pure
power law.  This $\tau$ as a function of $T$ will
have a minimum which we label $\tau_{min}$. Since at $T_c$, the fit
is strictly a power law, one can accept that temperature as $T_c$ where
the $\chi^2$ of the fit is minimum. However, we will determine $T_c$ using
the method described in step 3.

(2) Let $z=a^{\sigma}(T-T_c)$; $f(z)$ has a maximum at some value of
$z=\tilde z$: $f_{max}=f(\tilde z).$
For each mass number $a$ the yield $<n_a>$ as a function of temperature 
has a maximum at some value of temperature, $T_{max}(a)$.  
At this temperature $<n_a>(max)=a^{-\tau}f_{max}$ where $f_{max}$
is a constant independent of $a$.  This relationship allows us
to choose ``best'' values for $\tau$ and $f_{max}$.

(3) The value of $\tau$ found from step (2) is higher than $\tau_{min}$
found at step (1).  This means that if we look for $T$ appropriate
for $\tau$, two values of $T$ are available from the graph at step (1).
The lower value is to be chosen as the value of $T_c$.  The scaling
property (see steps (4) and (5)) is badly violated for the other choice.

(4) Now that we know $T_c$ and $T_{max}(a)$, the temperaure at which
the yield of composite $a$ is maximised, we find, by least squares fit
the ``best'' value of $\sigma$ from the condition 
$a^{\sigma}(T_{max}(a)-T_c)=const.$ for all $a$.

(5) The scaling law can now be tested by plotting $<n_a>a^{\tau}$
{\it vs.} $a^{\sigma}(T-T_c)$.  Plots for all $a$'s should fall on the
same graph.

In following steps (1) to (5) the range of $a$ is to be chosen 
judiciously.  It can not be very small (since Eq.(1.1) applies 
to ``large'' $a$'s).  But $a$ should also be truncated on the
high side (significantly smaller than the size of the dissociating system). 

\section{Results for one kind of particles}

We present our results in Figs.1 to 3.  The sizes of the systems are
taken to be $A=174$ and $A=240$.  The upper pannels of the figures 
use the freeze-out volume $V=3V_0$ and the lower pannels use $V=4V_0$.  
Both are shown here for completeness.  It will suffice here to discuss
only the cases with $V=3V_0$. 

Fig. 1 shows $\tau$ {\it vs.} $T$ drawn according to step (1) of section
3.  The dotted line is the value of $\tau$ deduced from step (2).
This cuts the curve of step (1) at two temperatures (step (3)).  
The lower value of the temperature is taken as $T_c$.  In the same
Fig. we also plot $C_v/A$ as a function of $T$.  The peak of this curve
corresponds to the first order phase transition, called
``boiling'' temperature in \cite{Dasgupta}.  It is remarkable that
the $T_c$ of Eq.(1.1) is very close to ``boiling temperature''in this 
particular example.  In this respect, the model is similar to the
behaviour of the lattice gas model where parametrisation Eq.(1.1)
works on the lower side of the critical density also provided $T_c$
is replaced by $T_b$, the temperature of first order phase transition
\cite {Pan2,Gulminelli1} at the given density.  In Fig.1 we have also plotted
the value of $\chi^2$ as a function $T$ (step 1).  In Fig.2 we plot
$ln<n_a>$ {\it vs.} $ln a$.  Two graphs are shown for each disintegrating
systems.  The graph with higher values of $<n_a>$ (shown as diamonds) follows 
from step (2) of section 3.  These are the maximum values of $<n_a>$ for
each $a$ obtained at corressponding temperatures $T_{max}(a)$.
The lower values of $<n_a>$ (shown as stars) are
all at the same temperature, namely at $T_c$ which, for example, is
6.32 MeV for $A$=174 and is 6.54 MeV for $A$=240.  This is how $\tau$
would be estimated from experiments \cite{Ogilvie}.  The crucial testing
of the scaling law is shown in Fig.3. where we plot $<n_a>a^{\tau}$
{\it vs.} $a^{\sigma}(T-T_c)$.  For the range of $a$ chosen (10 to 40)
the results nearly fall on the same graph.  Since one does not know
{\it a priori} how much error is due to finite particle number
of the disintegrating system, one might be tempted to to conclude
that the fit to Eq. (1.1) is adequate.  The parameters $\tau,\sigma$ from
best fits are 2.72, 1.06 respectively
for $A$=174 and 2.78, 1.23 respectively for $A$=240.
The deduced $T_c$ are 6.32 MeV and 6.54 MeV which are very different
from the critical temperature of 18 MeV for the model but
compare remarkably
well with the temperatures where the specific heats peak and which
correspond to first order phase transition temperatures at the
given densities.

Here we want to comment that, as seen from Fig. 1, $\chi^2$ has
a minimum at a temperature very close to $T_c$. So one may conclude
that the methods of determining $T_c$ using step 1 or 3 yield almost
the same result.

These then are the two salient features: (1) Numerical fits of Eq.(1.1)
are surprisingly close and (2) interpreting $T_c$ as the critical temperature
is wrong although the deduced $T_c$ does correspond to a phase transition
temperature.  We briefly look at how these features hold 
for a more realistic model with isotopic spin.

\section{Models of two kinds of particles}

We considered the following simplified version.  Protons and neutrons are
elementary particles.  For deuteron, triton, $^3$He and $^4$He, experimental
binding energies are used but no excited states are included. 
For masses greater than 4 a semi-empirical formula is used.
Composites of $n$ neutrons and $z$ protons
have binding energies $W(n+z)-\sigma (n+z)^{2/3}-a_s\frac {(n-z)^2}
{(n+z)}-a_z\frac{z^2}{(n+z)^{1/3}}$.  Here $W$=15.8 MeV, $\sigma$=18 MeV
is the surface tension term (taken here as temperature independent), 
$a_s=23.5$MeV is the symmetry energy term and $a_z$=.72 MeV is the 
Coulomb energy term.  For composites $>$4 we also include excited states
in the Fermi-gas model (as in section II).  We also incorporate Coulomb
interaction between composites in the Wigner-Seitz approximation \cite{Bo}.

It is however also instructive to consider two kinds of particles with
the surface tension and symmetry energy term but with the Coulomb energy 
switched off.  Such matters, in the thermodynamic limit can be self-bound
at zero temperature.
With Coulomb included however matter would fall apart in the
thermodynamic limit even at zero temperature.  We show first the results 
for finite
systems with surface and symmetry energy terms but no Coulomb.  For
brevity we will show results only for the choice $V=3V_0$.  For two
kinds of particles, Eq.(1.1) can be used in more than one way.  The
subscript $a$ in Eq.(1.1) can stand for the mass number $a=n+z$.
Then Eq.(1.1) gives the distribution of particles of given mass
number, irrespective of $n$ or $z$. The lower panels of Fig. 4 to 6
use that.
Or, as is more suitable for experiments, $a$ can stand for $z$, the
charge, irrespective of the neutron number $n$. The upper panels of
Figs. 4 to 6 use that.  We notice that the deduced value of $T_c$ continues 
to be close to $T_b$ where the specific heat maximises.
 
For brevity, with Coulomb included, we show the results for $V=3V_0$ only.
Distribution of particles with a given charge, irrespective of the neutron
number $n$ are shown in Figs. 7 to 9.  The most noticeable difference
with the ``no Coulomb'' cases is that the deduced $T_c$ is now farther
removed from the temperature at which the specific heat maximises.  Coulomb
energy introduces rather significant changes to the caloric curves.
This was noticed before \cite{Bhatta}.  For example, without the
Coulomb, as the particle number increases, the peak of the specific heat becomes
narrower and the temperature at which the maximum is obtained slightly
shifts to higher value (see Fig. 4). The narrowing and the 
shifting of the peak seems
to be almost exactly compensated by the long range repulsive Coulomb
force as more and more charged particles are added
(Fig. 4 here and also Fig.(6) in \cite{Bhatta} where this is
discussed in greater detail).  Properties of nuclear matter but
where protons carry charges, require further studies.


\section{Fit to a Droplet Model}


We will now try to fit the predicted yields given by the model of section II
(one kind of particles) with a well-known droplet model\cite{Fisher}. 
An early application of the 
model, to heavy-ion collisions, can be found in \cite{Goodman}.  The model
has been revived recently \cite{Elliott}. 

In Fisher's model, the condensation of a real gas into large drops(clusters) of
liquid is modeled. This shares various similar features with 
multifragmentation models. For instance, the potential energy of large clusters
consists entirely of a bulk term and a term associated with the loss of
binding energy at the surface. There is no Fermi energy term, as 
the molecules inside the cluster are assumed to be Boltzmann distributed. 
The entropy of large clusters is, however, more
complicated. As clusters become large, the dominant effect may be 
ascribed, once
again, to a bulk term, and the remainder to a surface term. It was pointed out
that liquid clusters may not be restricted to spherical shapes, 
as is the case in most multifragmentation models. This prohibits the use of one form of 
surface area to parametrize the surface contribution. In reference 
\cite{Fisher} it is argued that, at low temperatures, the most important
configurations will be compact and globular. Their surface areas $\bar{s}$ are 
not much greater
than the minimum possible, and are assumed to admit the asymptotic 
condition that,
\begin{eqnarray}    
\bar{s}(k,\B) / k \ra 0 \;\;\;\;\;\;\;\; &\mbox{  as  }& 
\;\;\;\;\;\;\;\;  k \ra \infty \;\;\;\;\;\;\;\; \mbox { and } \nonumber \\
\bar{s}(k,\B) / \log{k} \ra \infty  \;\;\;\;\;\;\;\; 
&\mbox{  as  }& \;\;\;\;\;\;\;\; k \ra \infty. \label{surfcond}
\end{eqnarray}
Where, $k$ is the number of molecules occupying the cluster. If, 
for finite clusters,
one introduces this surface area to calculate various surface contributions, 
one
must introduce a correction term which varies as $\tau\log{k}$. The sign and
magnitude of $\tau$ is estimated from various other considerations involving
other models.

One may thus, very generally, obtain the mean number of clusters of size $k$ as  
\begin{eqnarray}
<n_k>=Ck^{-\tau}\exp((\mu_g-\mu_l)k/T+c_2k^{2/3}/T)
\end{eqnarray}
Here both $\mu_g$ and $\mu_l$ are functions of $T$.  At coexistence and
also at the critical temperature, they become equal to one another.
Also $c_2$ is a function of temperature
and at $T_c$ the coefficient $c_2$ goes to
zero.  Since above $T_c$ there is no distinction between the liquid and
the gas phase, one can not speak of droplets.  Thus the theory only applies
to $T<T_c$.  As such the formulation is more limited than the model of
Eq. (1.1) which applies to both sides of $T_c$.  The following fit was
tried.  We set $\tau=2$.  let $\alpha=(\mu_g-\mu_l)/T$, $\gamma=c_2/T$.
We fit the calculated $<n_k>$ to $Ck^{-2}\exp(\alpha k+\gamma k^{2/3})$
at different temperatures where $\alpha,\gamma$ values at each temperature
are varied for best fit.  The values of $\alpha, \gamma$ as
a function of temperature is shown in Fig. 10 where we also show rather
remarkable fit with the values of $<n_k>$ obtained from the model
of section II.  The values of $\alpha$ and $\gamma$ both go to zero
near temperature $T=6.5$ MeV suggesting that the critical temperature
is 6.5 MeV.


\section{Relationship between different models}


In this section a grand canonical approximation to the model of section II is
constructed. As in the previous section, this will be concentrated in the region
below the boiling temperature. A different parametrization of the output from
the exact canonical calculation will be obtained. Though this parametrization
has not been as throughly investigated as that of the previous section, it will
serve to provide a qualitative understanding of the behaviour of the yields
below the boiling temperature.

The yields(Fig.(\ref{nk_vs_k}), Fig.(\ref{fits})) 
are obtained from the exact expressions Eq.(\ref{canyld}). 
For the present analysis they are adequately approximated by  Eq.(\ref{grand}).
The parametrization offered by 
Eq.(\ref{grand}) is of the form

\begin{eqnarray}
\lc n_k \rc = \T{C} k^{3/2} \exp ( \T{\ga} k + \T{\g} k^{2/3} ) \label{grandfit}
\end{eqnarray}

\nn
The above is different from the parametrization of the droplet model where 
$\tau = 2$. Though the expressions look similar, the fit parameters 
$\T{C},\T{\ga},\T{\g}$ will assume values different from those of 
$C, \ga, \g$. The interpretation of $\T{\ga}$ is also quite precise in this
approximation.

When using Eq.(\ref{grand}) to estimate the result of a canonical calculation, 
the free
parameter is the chemical potential $\mu$. It is usually determined by 
imposing that the
model correctly reproduce the total number of particles 
composing the system(i.e., $\sum_k  \lc n_k \rc k = A$ ).
This is a complicated problem in general. A clue may be obtained by 
observing the behaviour
of $\mu$ as obtained from the canonically calculated Helmholtz free energy
$F=-T\log{Q_A}$.
A plot of $\mu$ obtained thus is plotted in Fig.({\ref{mu}})(data
points). The
behaviour of this $\mu$ may be estimated by the following simple argument. 
Far below the
boiling temperature the system exists mostly as one large cluster 
and a few small ones(see Fig. (\ref{nk_vs_k})). 
The large cluster is considered as the liquid 
state. Far above the 
boiling temperature the
system exists mostly as many small clusters: 
this is considered as the gas phase. 
The chemical potential of either system may 
be estimated by keeping the system 
in contact
with a heat reservoir, adding one particle to 
the system, and noting the 
change in free
energy, i.e.,
$\mu = \Big[ \Delta F \Big]_{V,T} = F(T,V,A+1) - F(T,V,A)$.
On entering the system, the new particle, may apriori 
attach itself to any of the existing
clusters, or simply thermalize as a monomer in the system. 
It will attach
itself to the cluster that minimises the free energy at that 
temperature and density.
There may be more than one unique choice. 
The resulting change in energy and entropy of the system may be 
decomposed as the sum
of two parts: a kinetic part($\D E_{kin}, \D S_{kin}$), 
to do with the cluster's motion in the environment; and an
internal part($\D E_{in}, \D S_{in}$), 
to do with the internal motion of the particles constituting the cluster.
If the volume is large, one may assume that the clusters form 
an almost ideal gas. 
In this case the average kinetic energy of a cluster of size 
$k$ is $(3/2)T$. It does not
depend on $k$ and thus $\D E_{kin} = 0$. The change in internal 
energy $\D E_{in}$ may be
estimated simply as, 

\begin{eqnarray}
\D E_{in} = -W + \frac{T^2}{\e} + \sg (T)\big[ (k+1)^{2/3} - k^{2/3} \Big].
\end{eqnarray}

The change in internal entropy is given simply as $ \D S_{in} = 2T/\e $. 
The kinetic 
entropy of an ideal gas of $n_k$ clusters is given 
as(see Ref. \cite{zem}),

\begin{eqnarray}
S_{kin} = n_k \Bigg[ \frac{3}{2}\log{T} + \log{V} + 
\frac{3}{2}\log { \frac{2\pi m}{h^2} } + \frac{3}{2}\log{k} + \frac{3}{2} \Bigg] -\log{n_k!}.
\end{eqnarray}

\noindent
In most cases in nuclear fragmentation, $n_k$ lies between $0$ and $1$
(see Fig. (\ref{nk_vs_k})). Thus we may ignore the $n_k!$ term. Thus 
we get the total change in entropy for the addition of one particle to a 
cluster of size $k$ as,

\begin{eqnarray}
\D S = 2T/\e  + \frac{3}{2}\log \Big( 1+1/k \Big).
\end{eqnarray}

\noindent
As a result, the total change in free energy and hence $\mu$,
is given as 

\begin{eqnarray}
\mu = \D F = -W - \frac{T^2}{\e} + \sg (T) \Big[ (k+1)^{2/3} - k^{2/3} 
\Big] -
\frac{3}{2} T \log \Big( 1+1/k \Big).
\label{effmu}
\end{eqnarray}

We note that $\mu$ becomes progressively more negative with rising $k$. 
Thus the added particle prefers to attach to large clusters. In a fragmenting 
system under the boiling temperature, such a large fragment exists, of about half the size
of the system \cite{Dasgupta}. The new particle thus preferentially attaches to this
cluster. To illustrate this point more quantitatively, we calculate this $\mu$ for a system
with $A=240$ and a $V=4V_0$. We assume the largest cluster is of size $A/2$.
No doubt, this size falls gradually with rising temperature \cite{Dasgupta}, with the fall
becoming rapid near the boiling temperature. In Fig. (\ref{mu}) we plot the value of this
$\mu$ (solid line) assuming that the largest cluster remains of the same size throughout
the temperature range.
Above the boiling point the system exists mostly as small clusters, 
here we assume that the added
particle attaches itself to a cluster of size $k=2$. This $\mu$ is plotted as the
dot-dashed line in Fig. (\ref{mu}). 

We are interested in obtaining an approximate expression for $\lc n_k \rc$ underneath the
boiling temperature. Using the expression for $\mu$ as derived in Eq. (\ref{effmu}), we
obtain the expression for $\lc n_k \rc$ as

\begin{eqnarray}
\lc n_k \rc = e^{\B \mu k}\w_k &=& 
V (\frac{2\pi m T}{h^2})^{3/2} k^{3/2} \nonumber \\ 
& & \times \exp \Big[ \Big\{ \sg(T)
\Big( (k_{max} + 1)^{2/3} - k_{max}^{2/3} 
\Big) -\frac{3}{2}T \log (1 + 1/k_{max}) \Big\} k/T 
- \sg(T) k^{2/3}/T \Big] .
\end{eqnarray}

For most systems, in general, the behaviour of $k_{max}$ with $T$ and $V$ 
is difficult to estimate. However, we note that the above equation is 
precisely of the form of Eq. (\ref{grandfit}).
On fitting the data points obtained from Eq. (\ref{canyld}) 
we obtain the fit parameters as $\T{C}=2.73, \T{\ga}=0.36, \T{\g}=-3.11$ 
(note that, as in the droplet model, only range of $k$ between 10 to 40, is fitted).
A plot of the fit to the values of $n_k$ obtained from a system with
$A=240$, $T=5$MeV and $V=4V_0$ is shown in Fig. (\ref{fits}). Here 
the entire region from
$k=2$ to $240$ is plotted. Note that both fits 
coincide extremely well in the region of $k= 10$ to $40$.

\section{summary and discussion}
This investigation started out with the following question: 
suppose ``experimental data'' are given by the predictions of a 
theoretical model which,
we know, does not conform to Eq.(1.1) exactly.  Could we still
describe the ``data'' approximately with the formula?  And, if so,
what are the significances of the parameters $\tau,\sigma$ and $T_c$?
Subsequently we tried a fit with the droplet model and again found
that a very adequate fit can be found. 

We find that the scaling law is still approximately obeyed.
While we can not attach much significance to the extracted values of
$\tau$ or $\sigma$ (they are different from those given by the
percolation or Ising model, but not terribly so), $T_c$ seems to 
be a genuinely physical parameter, namely, it reflects the
first order phase transition temperature.  Coulomb effects tend
to somewhat spoil even this correspondence.

Many different fits can be obtained because by necessity the mass
number of the composites is limited on the lower as well as on the
higher side.  We have no control over that since the dissociating
systems are extremely finite.  This apparently makes even deciding
on the order of phase transition very difficult.  Fits to data
from calculations on percolation model have been made in the past
\cite {Elliott2}.  Since this is a model of continuous phase
transition one necessarily concludes that the phase transition
is continuous.  It is more appropriate to consider the
Lattice Gas Model \cite {Pan1} instead which does encompass the
Percolation model as a subset \cite {Subal}.  In the Lattice Gas  model, 
if it is assumed that that the freeze-out density is less than half
the normal density, then the fit of Eq. (1.1) would imply
that $T_c$ is indeed a first order phase transition temperature.
If, on the other hand, one assumed that the freeze-out density
is higher than half the normal density (we consider this highly
unlikely), then a fit to Eq. (1.1) would not imply any usual 
thermodynamic phase
transition \cite {Campi}.
If one depends on theories to decide on what order of phase transition to
expect, one is driven towards expecting a first order phase transition.
To date all the models which used a Hamiltonian 
\cite {Gulminelli1,Pan2,Bo,Dasgupta} suggest a first order phase
transition.

\section{Acknowledgments}
This work was supported in part by the Natural Sciences and Engineering 
Council of Canada and by {\it le Fonds pour la Formation de chercheurs
et l'Aide \`a la Recherche du Qu\'ebec}.

\begin{figure}
\caption{$\tau$, ${C_v}/A$ (left-hand scale) and $\chi^2$ (right-hand scale)
plotted against temperature in the model of one kind of particles. The different
panels are for different choices of $A$ and $V$.}
\end{figure}

\begin{figure}
\caption{$ln <n_a> vs. ln a$. The solid line is the best fit to $ln <n_a>$
at each $T_{max}(a)$ presented by diamonds. The dotted line joining stars 
represents the distribution at $T_c$.The diferent
panels are for different choices of $A$ and $V$.}
\end{figure}

\begin{figure}
\caption{The scaling behavior in the mass range $(10 \le a \le 40)$}
\end{figure}

\begin{figure}
\caption{Same as Fig. 1, but in a model of two kinds of particles. Coulomb
interaction has been switched off. The upper and lower panels are with
respect to charge and mass distributions.} 
\end{figure}

\begin{figure}
\caption{Upper panels: $ln <n_z> vs. ln z$, and lower panels: 
$ln <n_a> vs. ln a$; in the model of two kinds of particles with Coulomb
interaction switched off.}
\end{figure}

\begin{figure}
\caption{Scaling behavior in the charge range $(7 \le z \le 17)$ (upper
panels) and mass range $(10 \le a \le 40)$ (lower panels) in two kinds of 
particles model without Coulomb interaction.}
\end{figure}

\begin{figure}
\caption{$\tau_z$ and ${C_V}/A$ as a function of temperature in a
model of two kinds of particles, with Coulomb interaction included.} 
\end{figure}

\begin{figure}
\caption{Same as upper panels of Fig. 5, but with Coulomb interaction.}
\end{figure}

\begin{figure}
\caption{Same as upper panels of Fig. 6, but with Coulomb interaction.}
\end{figure}

\begin{figure}
\caption{The parameters of droplet model $\alpha$ and $\gamma$ as a function
of temperature for a system with $A=240$ and $V=4V_0$. 
The right pannels show the fit of the droplet model to the
yields obtained in the model of one kind of particles described in section II. 
On the graph one
can not distinguish between fitted points and the actual points from 
canonical calculations. }
\end{figure}

\begin{figure}
\caption{$<n_a>$ vs. $a$ on a logarithmic plot. The right pannel expands the
region $10 \le a \le 40$.} \label{nk_vs_k}
\end{figure}

\begin{figure}
\caption{ $\mu$ vs. $T$ for a system with $A=240$ and $V=4V_0$. 
Data points represent results from a canonical
calculation (see section II). Solid line represents $\mu$ for addition to the largest 
cluster, dot-dashed line is 
$\mu$ for addition to a small cluster (see section VII for details). }
\label{mu}
\end{figure}

\begin{figure}
\caption{ Fits to $<n_a>$ vs. $a$ from two different models: open circles are from an exact canonical
calculation; 
the solid line represents the fit by a droplet model; 
the dotted line represents the fit from the grand canonical
approximation. }
\label{fits}
\end{figure}

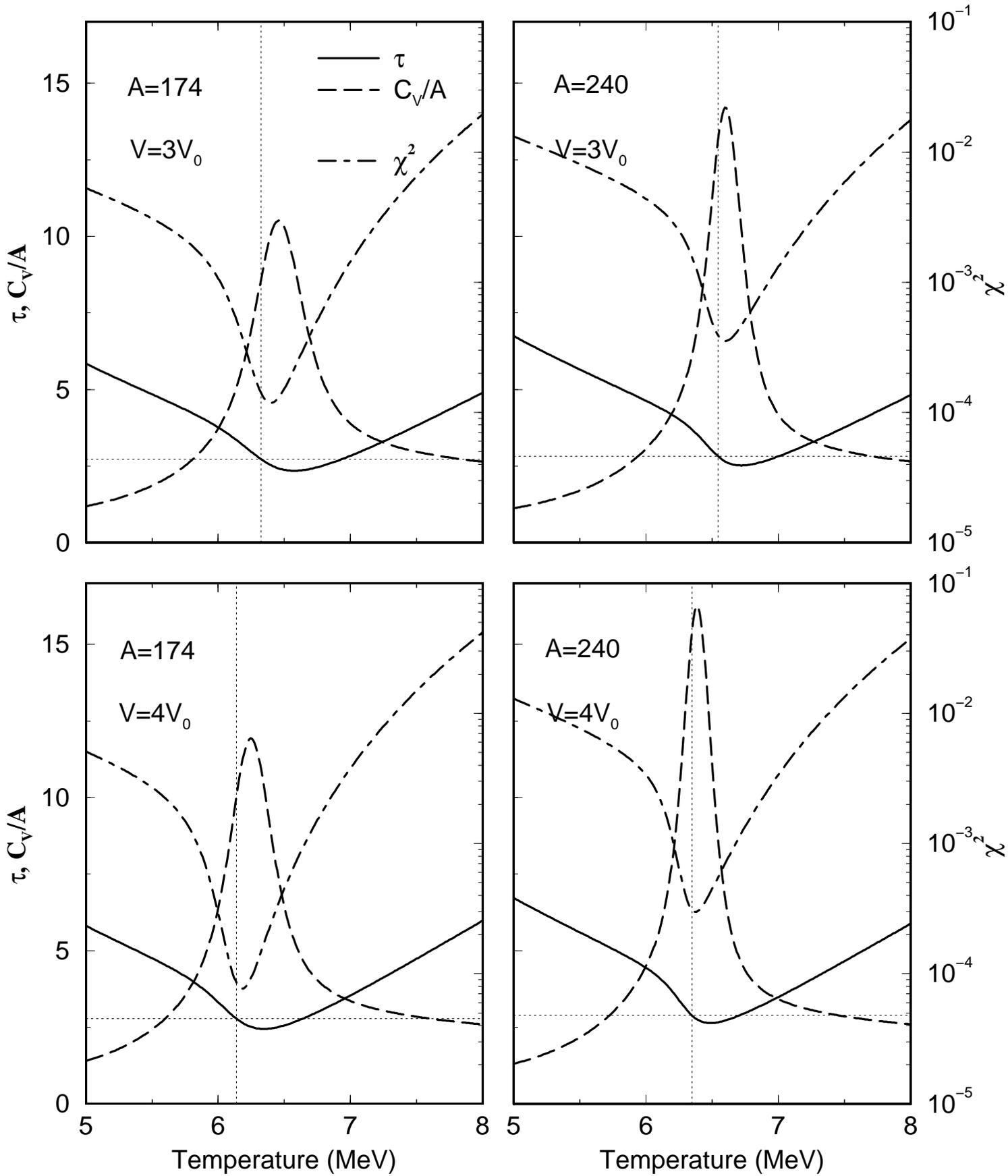
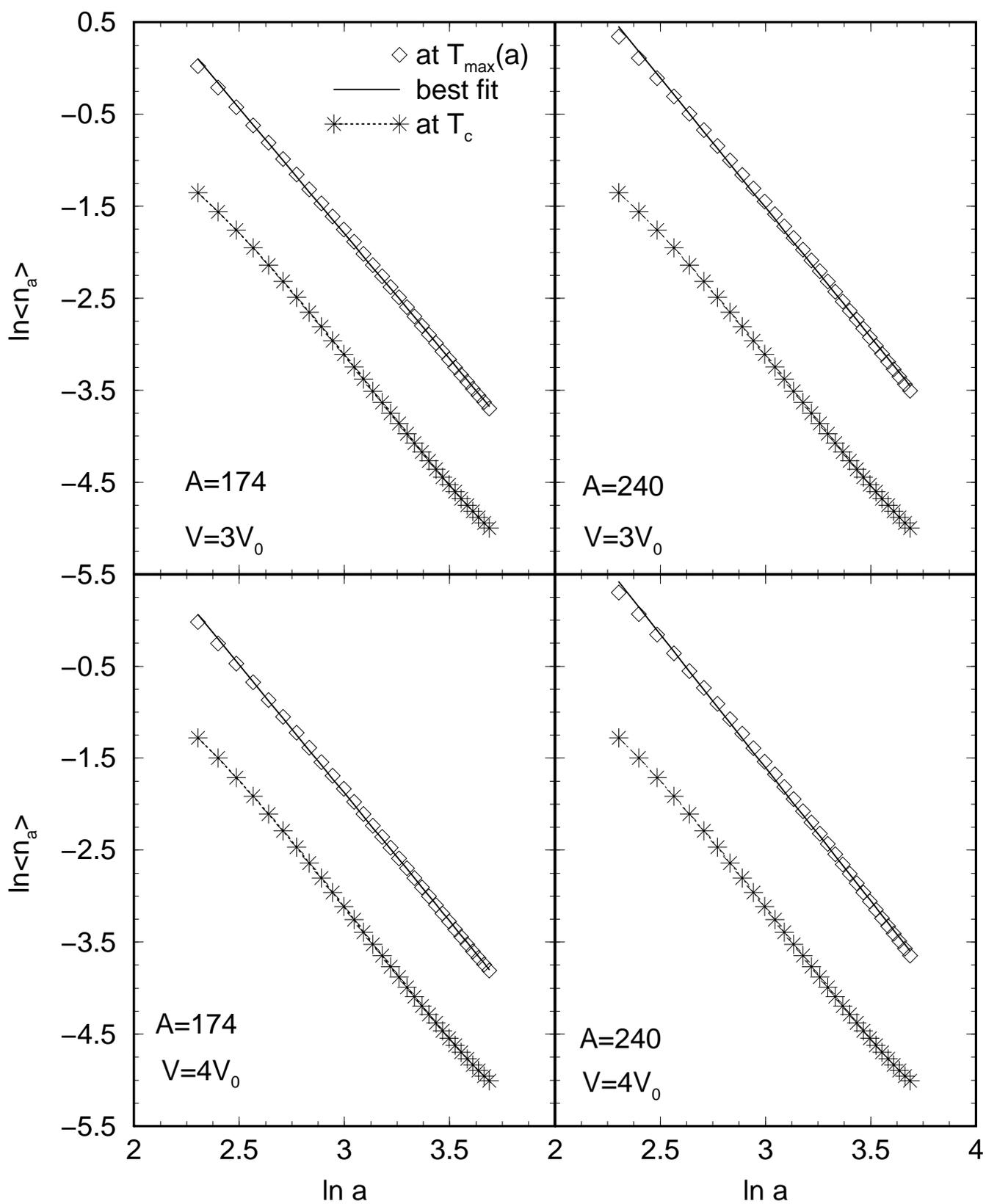
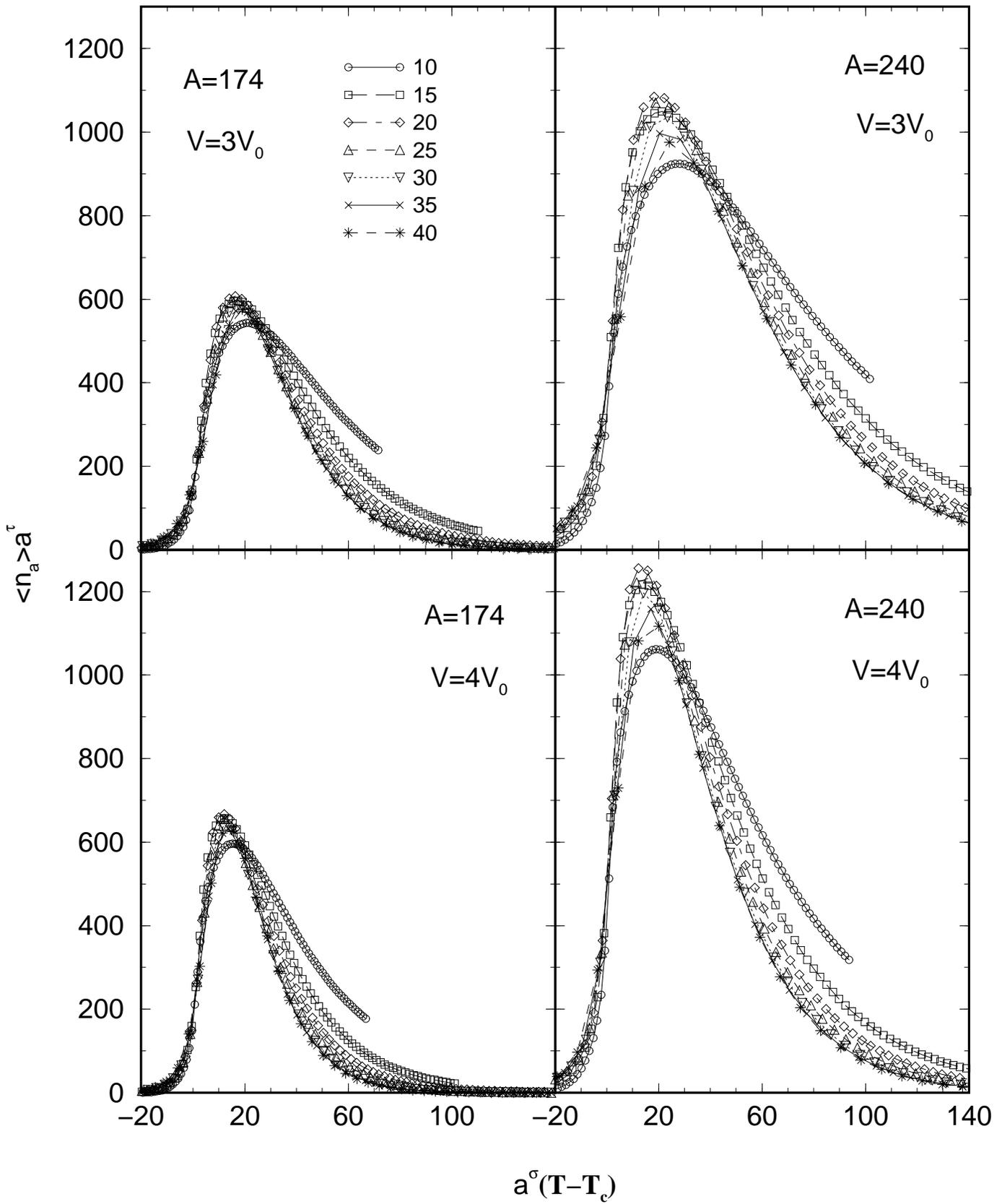
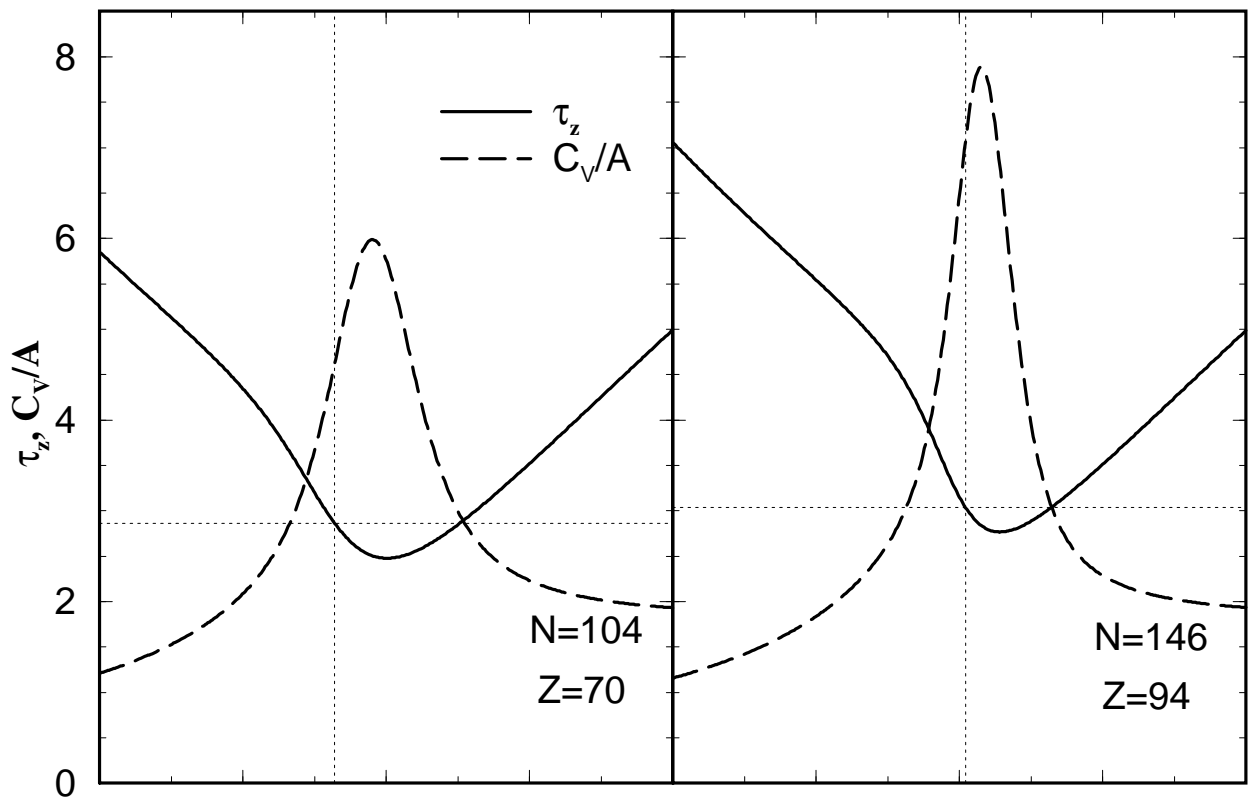
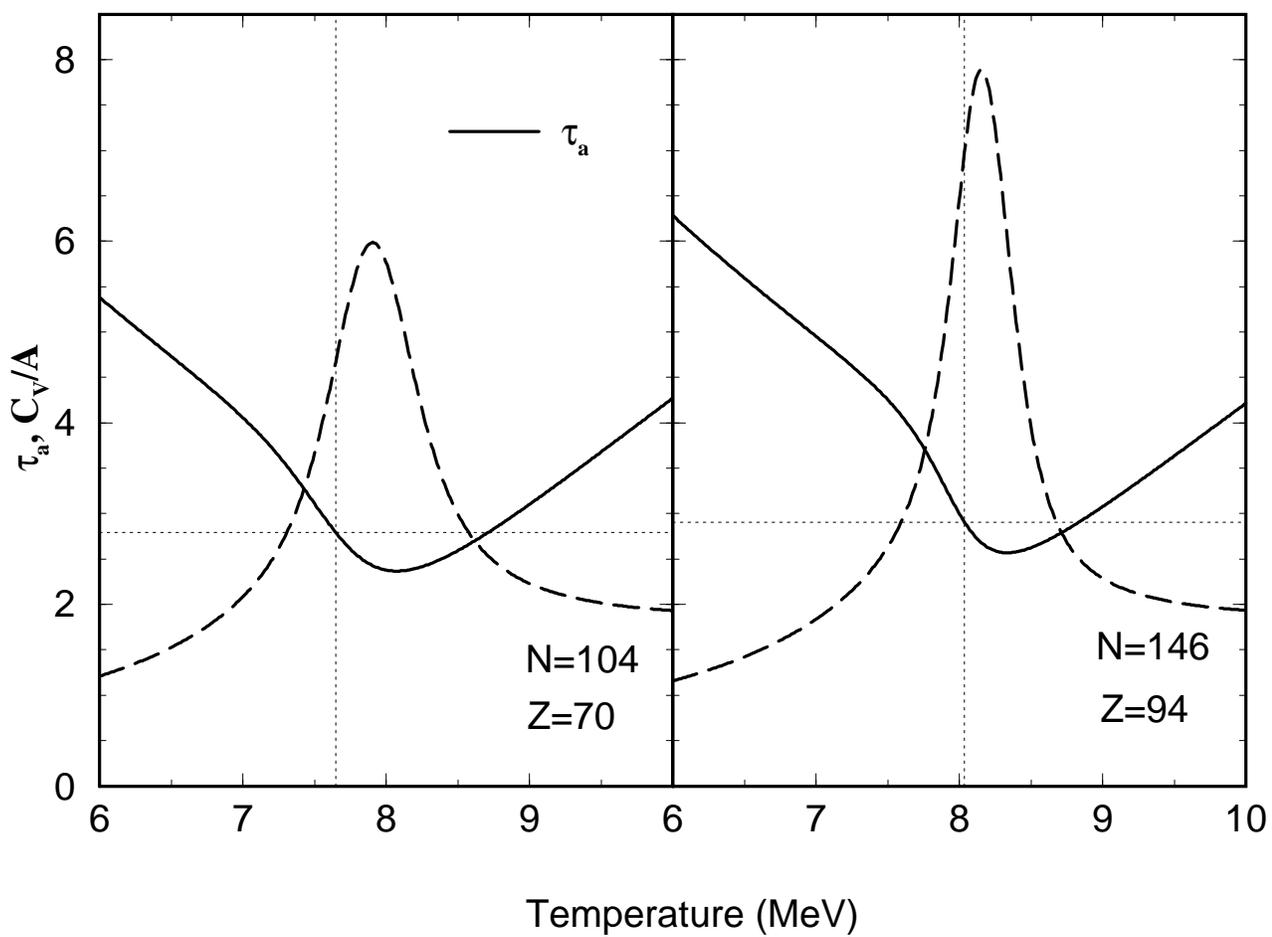
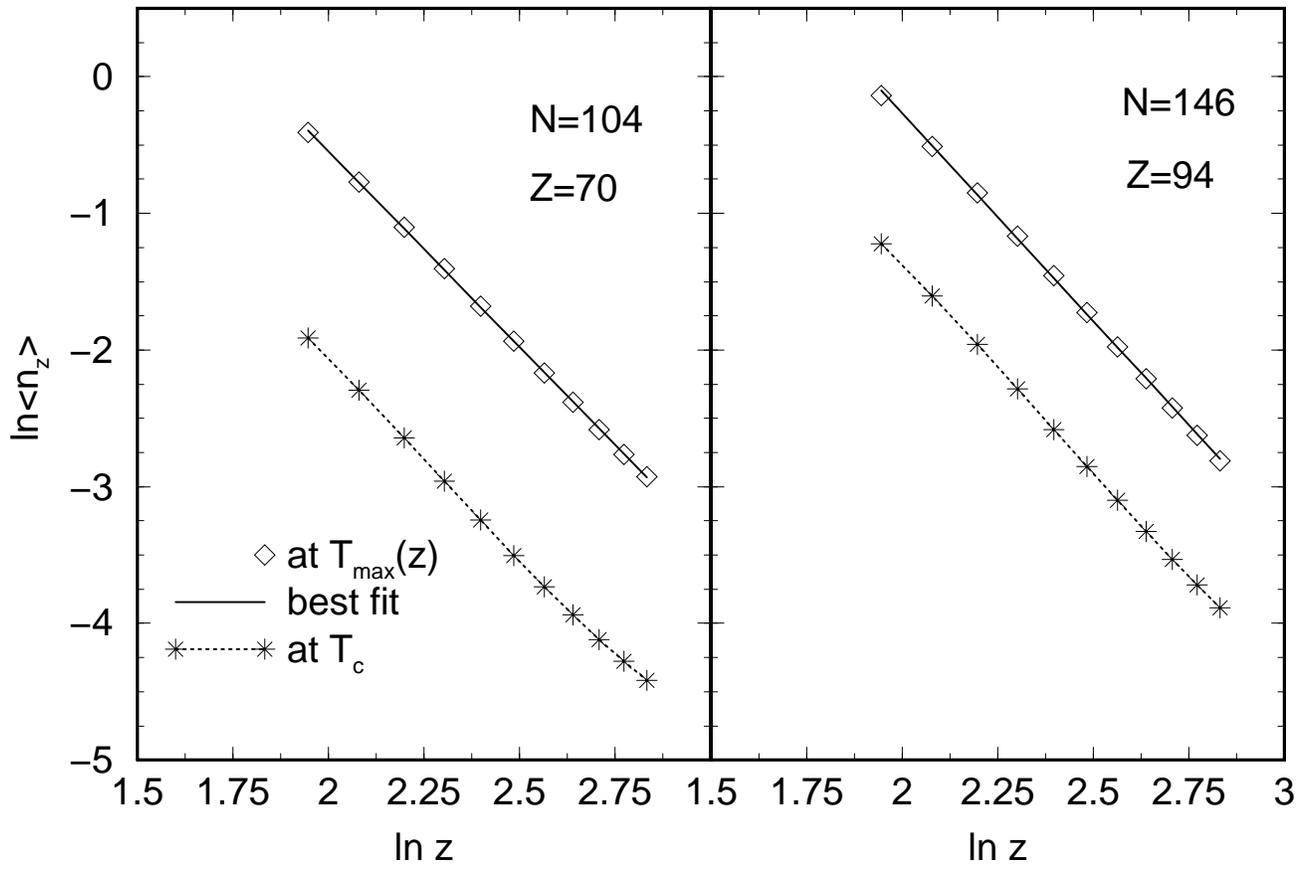
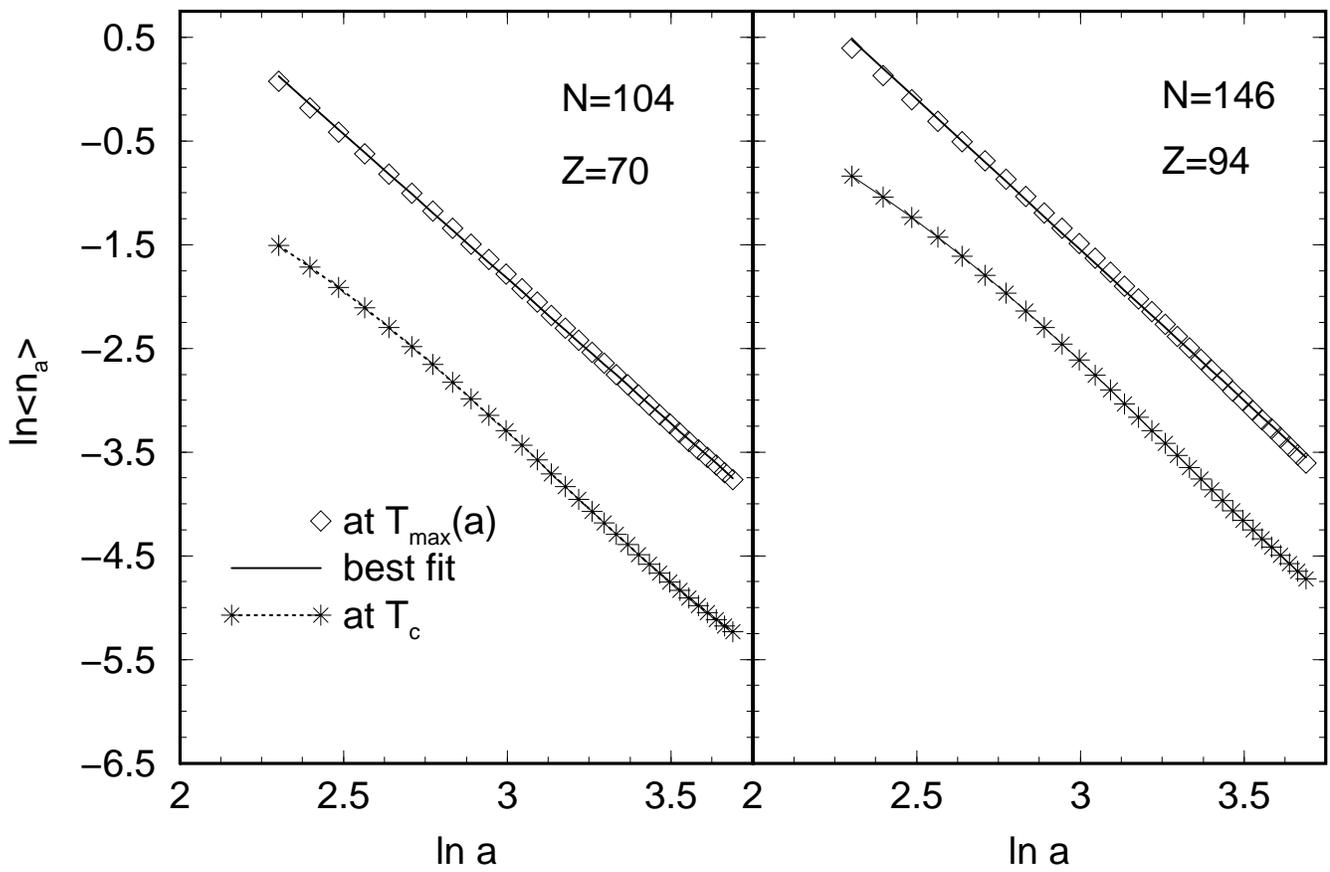
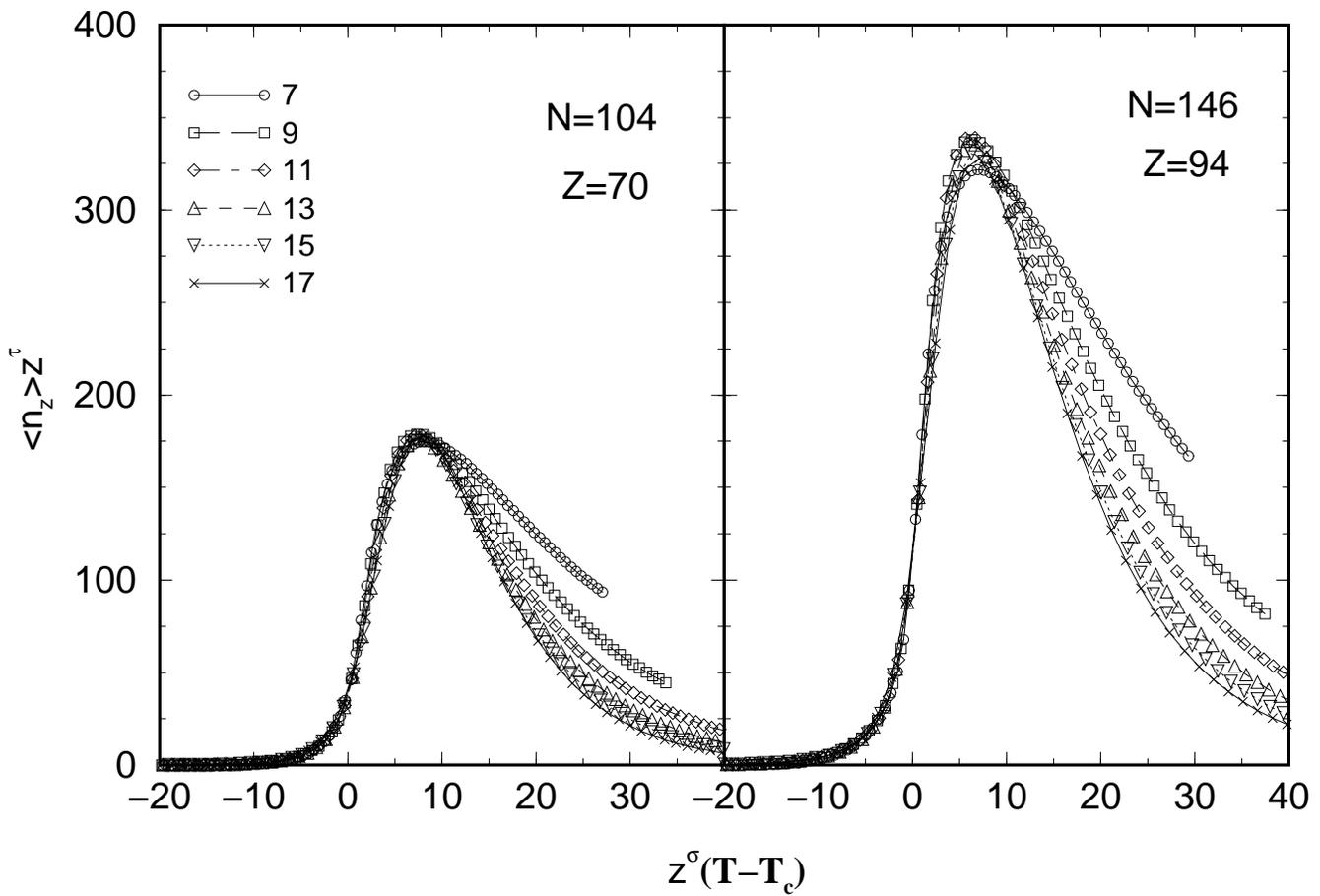
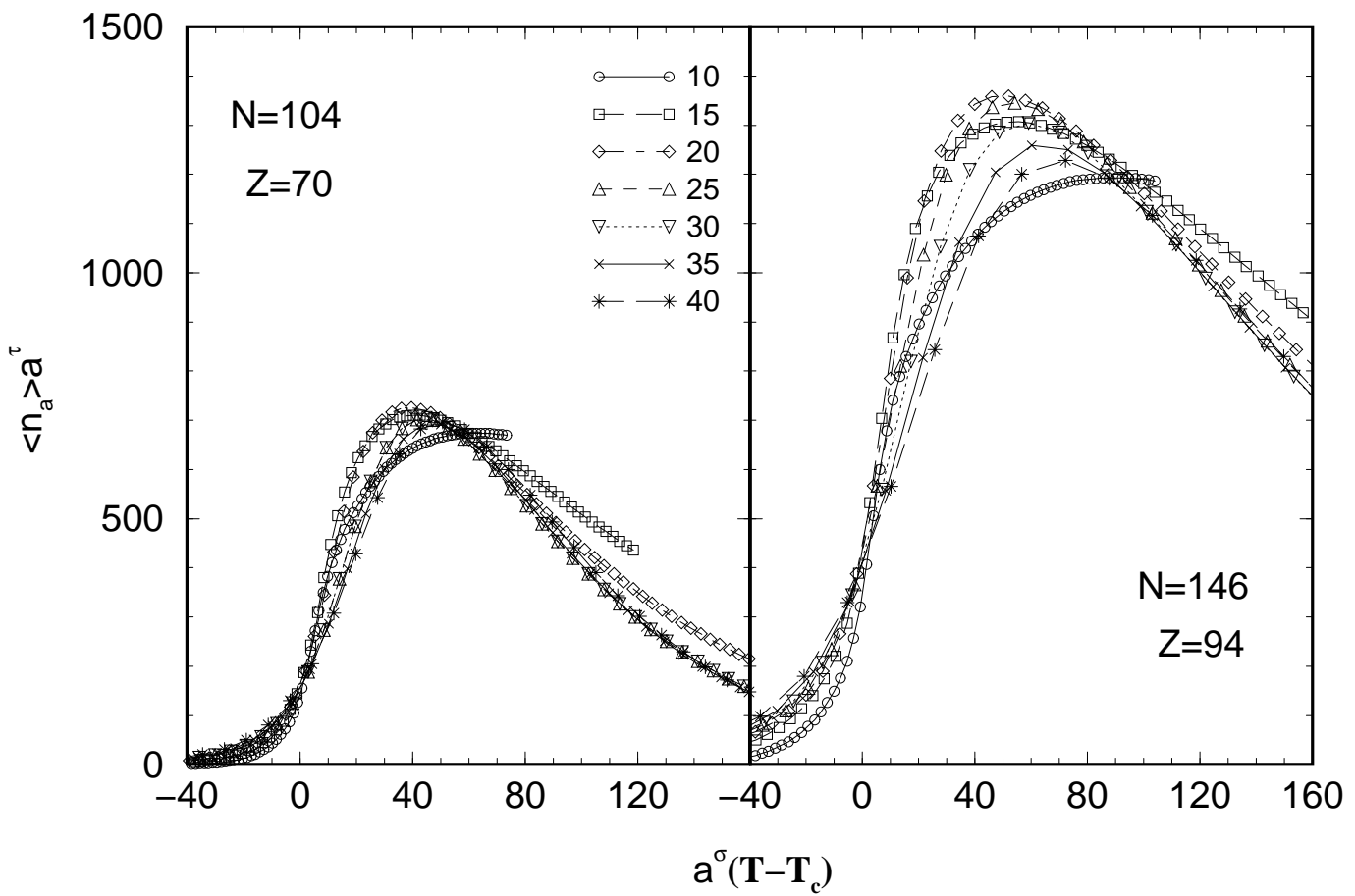
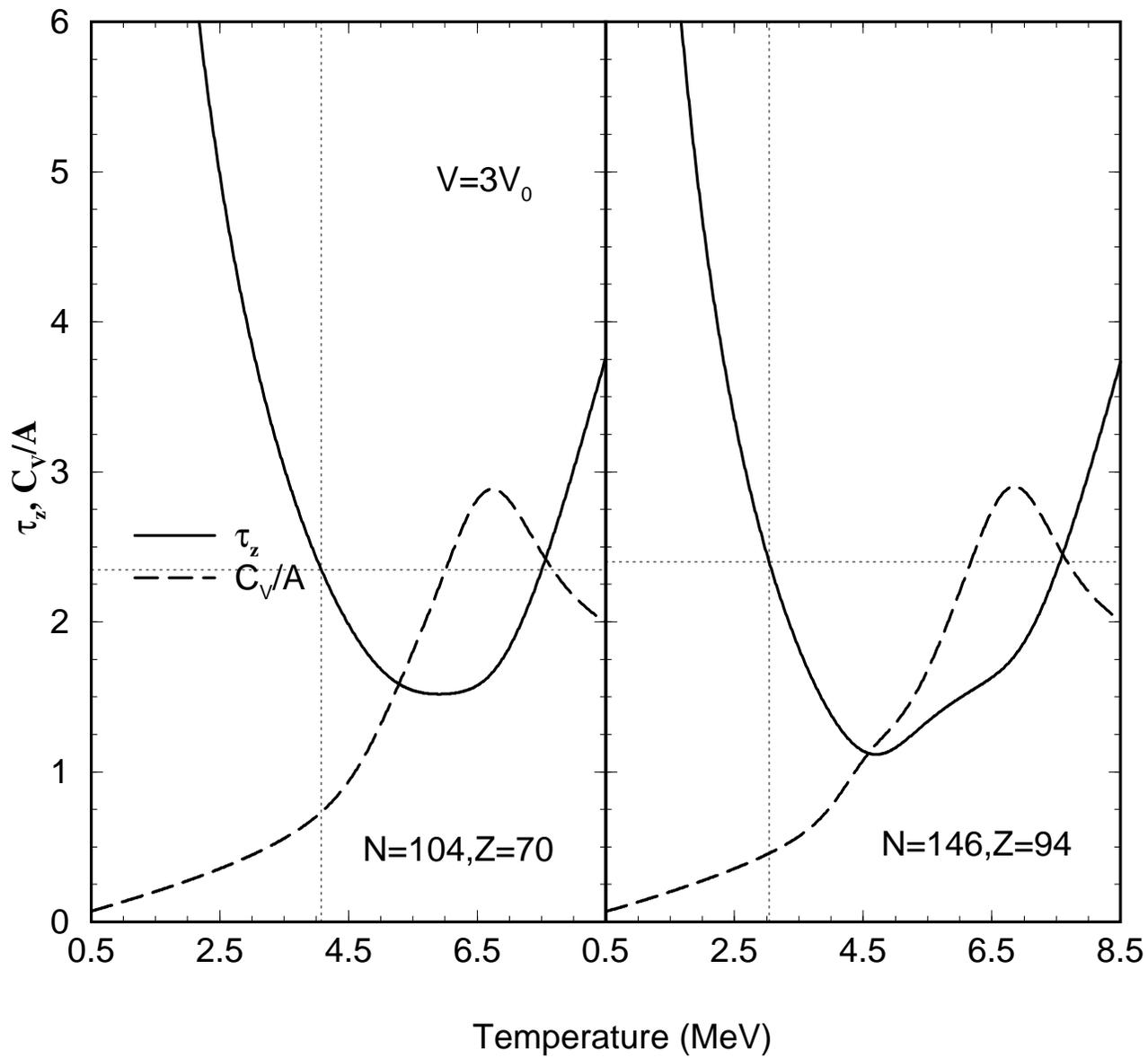
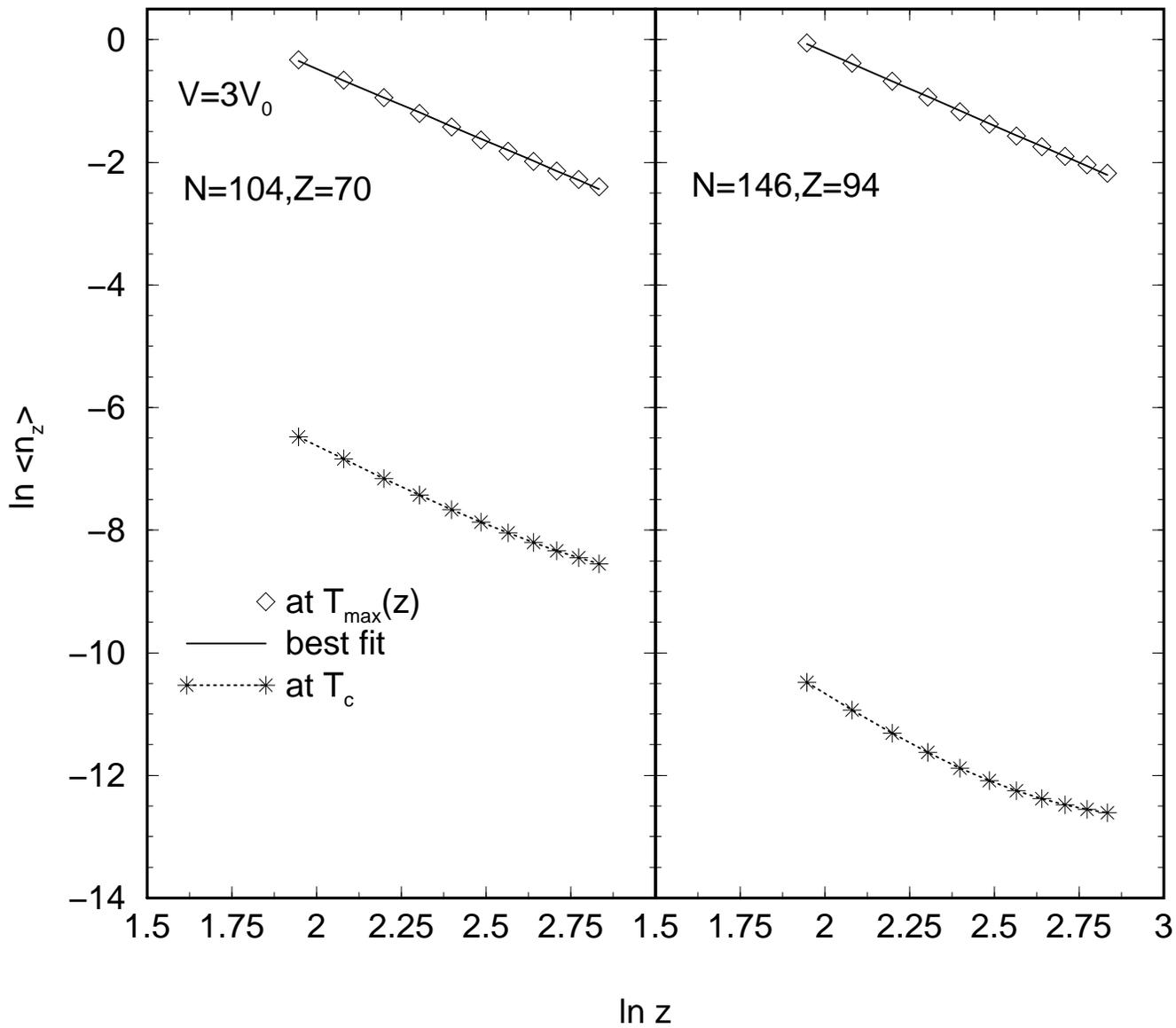
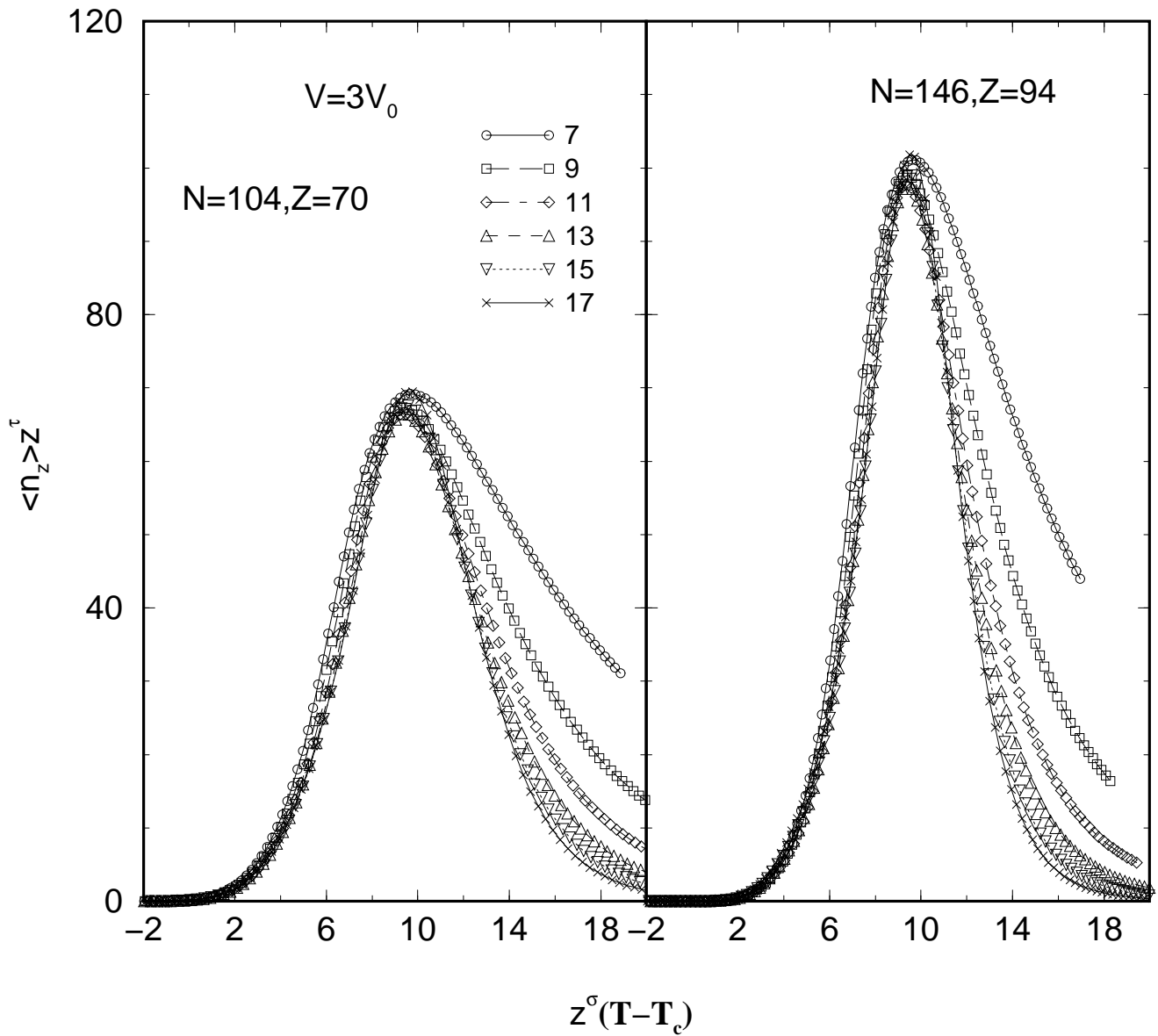
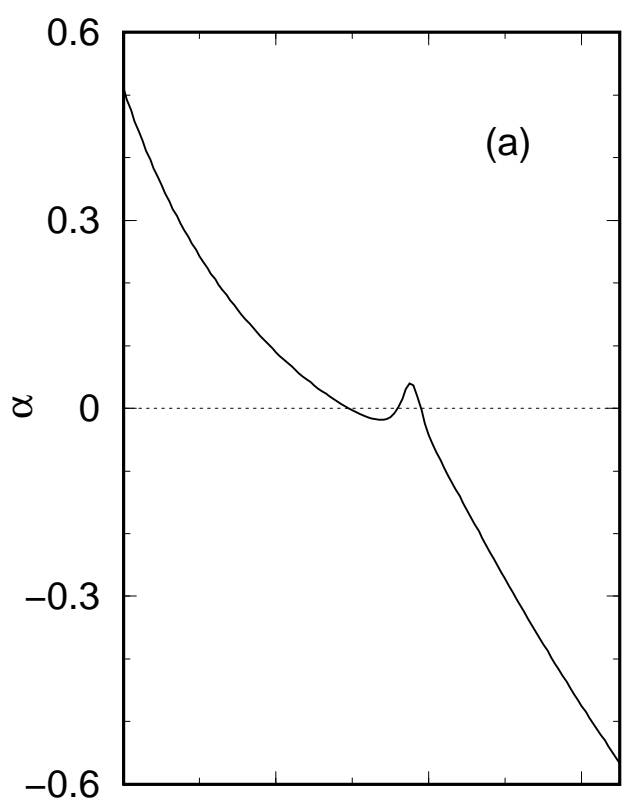
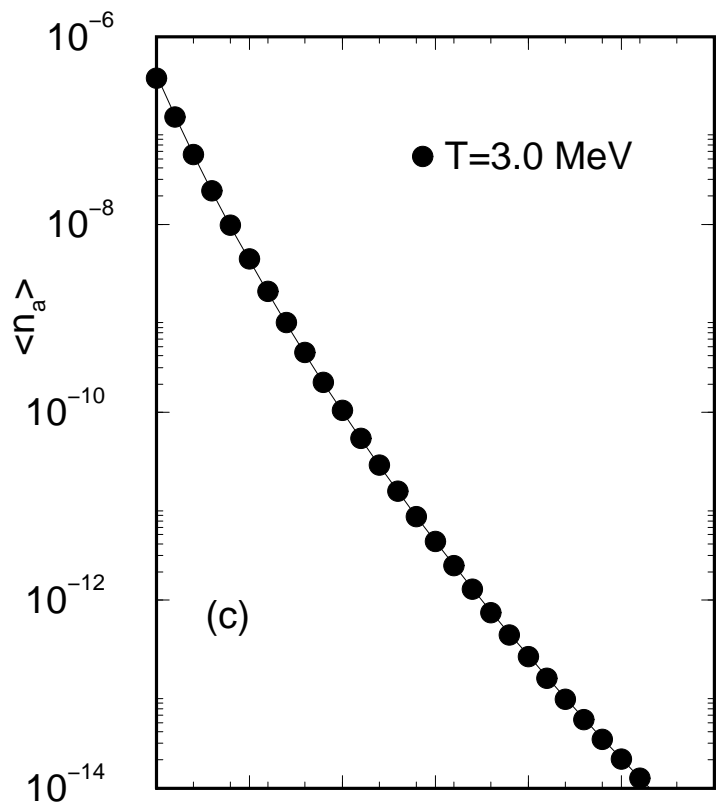
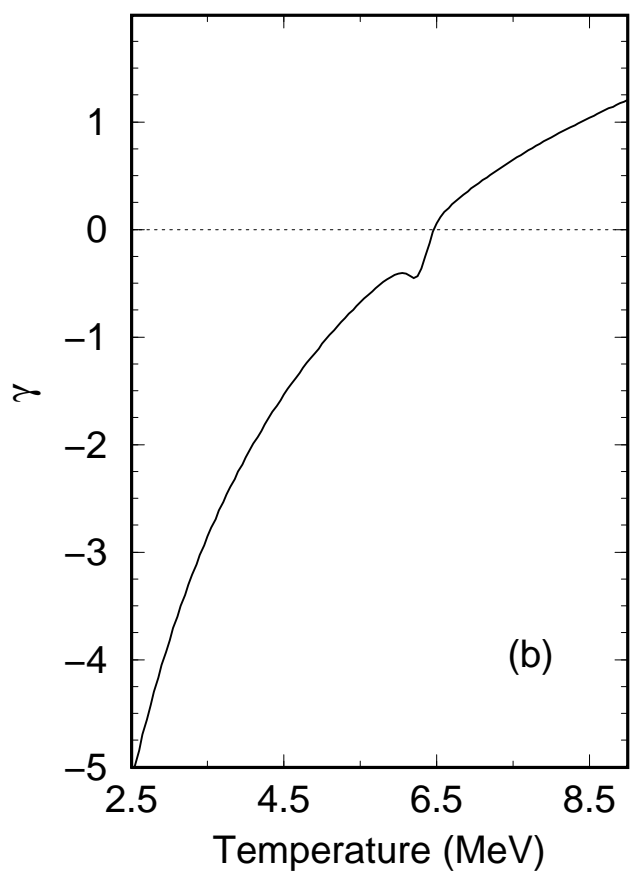
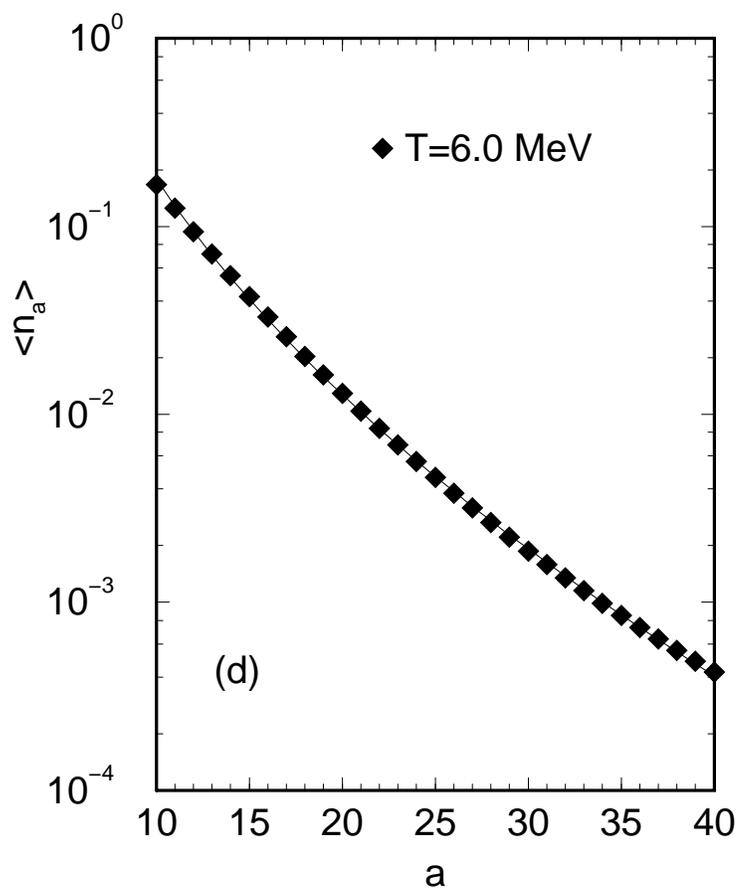
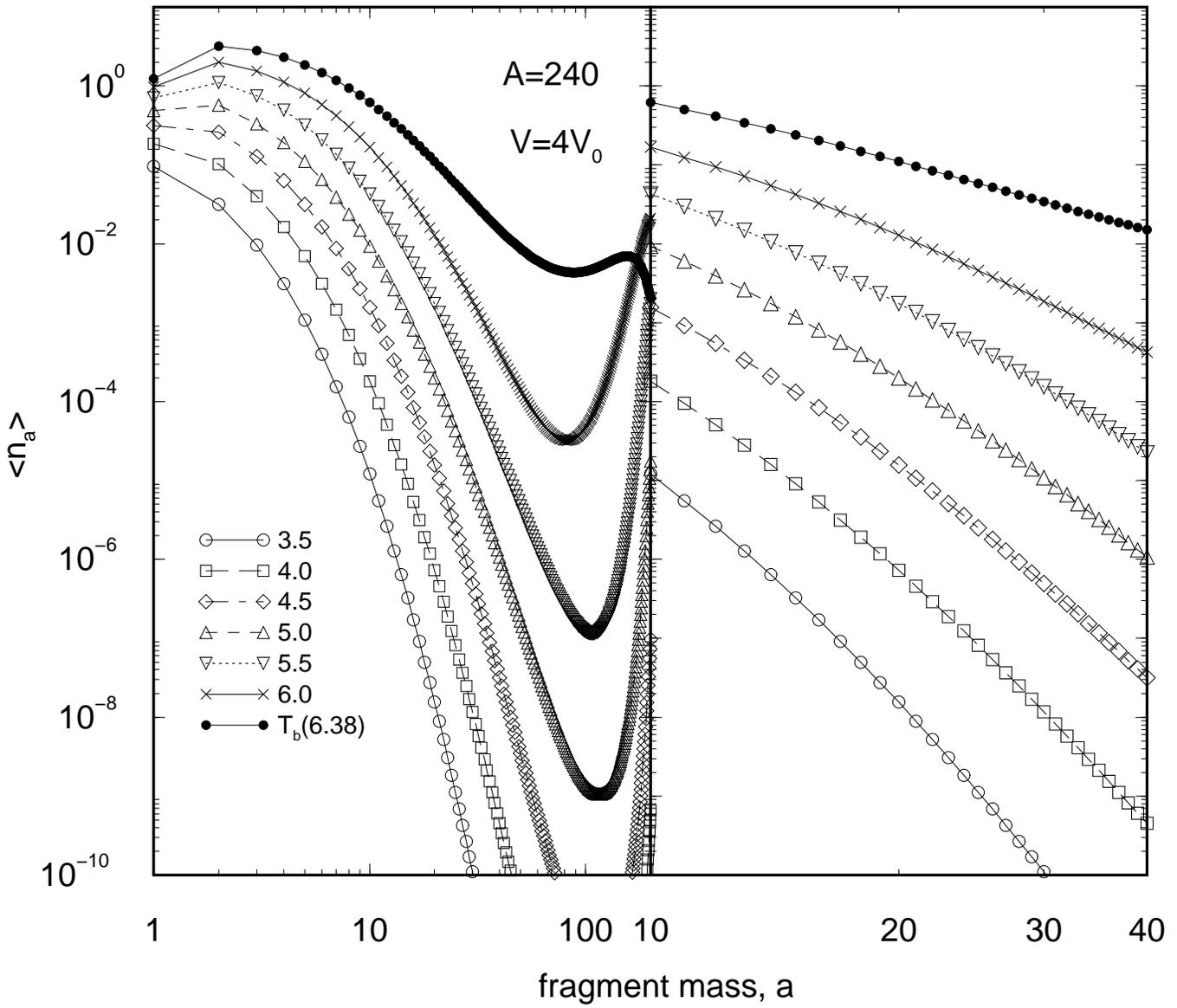
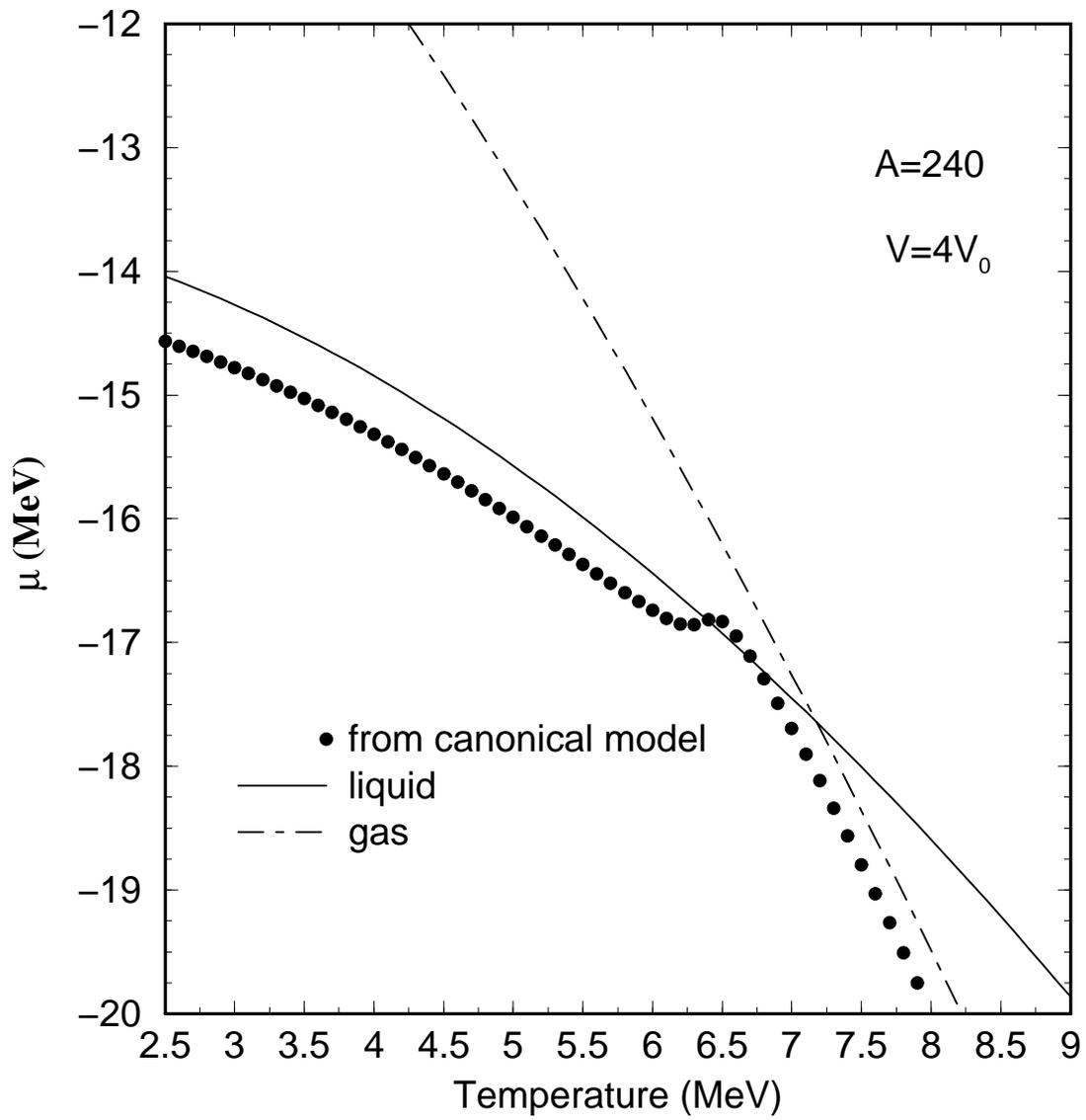
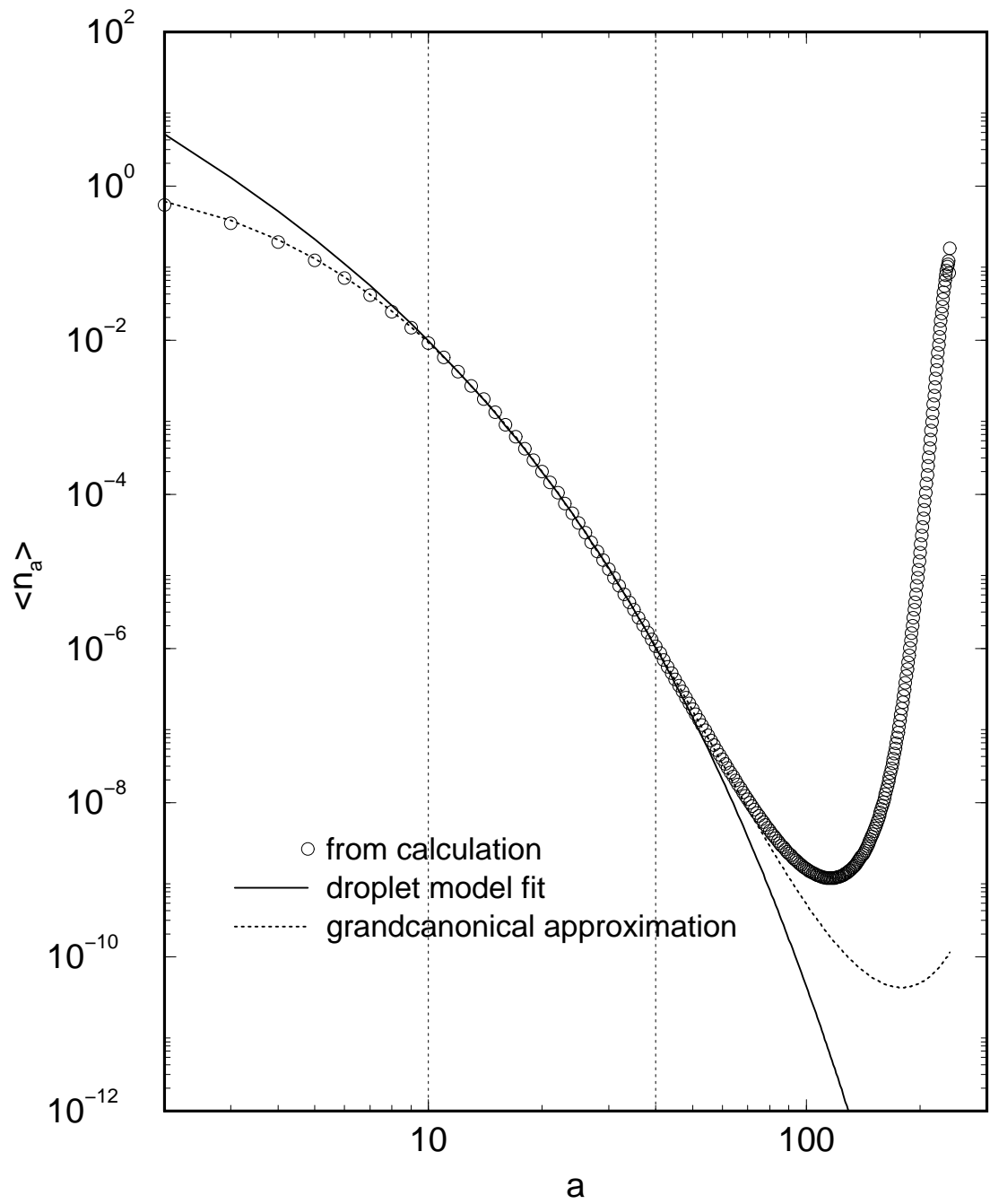
\end{document}